\documentclass{ieeeaccess_arxiv}
\usepackage{cite}
\usepackage{amsmath,amssymb,amsfonts}
\usepackage{algorithmic}
\usepackage{graphicx}
\usepackage{textcomp}
\usepackage{booktabs}
\usepackage[utf8]{inputenc}
\DeclareUnicodeCharacter{200B}{}
\usepackage[T1]{fontenc}
\usepackage{lmodern}
\usepackage{float}
\usepackage{adjustbox}
\usepackage{tabularx}
\usepackage{array}
\newcolumntype{L}{>{\raggedright\arraybackslash}X}
\usepackage{threeparttable}
\usepackage{pdflscape}

\usepackage[colorlinks=true, linkcolor=blue, citecolor=blue, urlcolor=blue]{hyperref}

\usepackage{bm}
\makeatletter
\AtBeginDocument{\DeclareMathVersion{bold}
\SetSymbolFont{operators}{bold}{T1}{times}{b}{n}
\SetSymbolFont{NewLetters}{bold}{T1}{times}{b}{it}
\SetMathAlphabet{\mathrm}{bold}{T1}{times}{b}{n}
\SetMathAlphabet{\mathit}{bold}{T1}{times}{b}{it}
\SetMathAlphabet{\mathbf}{bold}{T1}{times}{b}{n}
\SetMathAlphabet{\mathtt}{bold}{OT1}{pcr}{b}{n}
\SetSymbolFont{symbols}{bold}{OMS}{cmsy}{b}{n}
\renewcommand\boldmath{\@nomath\boldmath\mathversion{bold}}}
\makeatother

\def\BibTeX{{\rm B\kern-.05em{\sc i\kern-.025em b}\kern-.08em
    T\kern-.1667em\lower.7ex\hbox{E}\kern-.125emX}}

\begin{document}
\history{}
\doi{}
\title{Integrating Persuasion Theory into the Epidemiological
       Modelling of Health Misinformation Spread on Social Media}
\author{
  \uppercase{Mkululi Sikosana}\authorrefmark{1},
  \uppercase{Sean Maudsley-Barton}\authorrefmark{1},
  \uppercase{Oluwaseun Ajao}\authorrefmark{1}
}

\address[1]{Department of Computing and Mathematics,
            Manchester Metropolitan University,
            Manchester M15 6BH, United Kingdom}

\corresp{Corresponding author: Mkululi Sikosana
         (e-mail: mkululi.sikosana@stu.mmu.ac.uk).}

\tfootnote{This work did not receive any financial support.}

\markboth
  {Sikosana \headeretal: Epidemiological Modelling of Health Misinformation Spread on Social Media}
  {Sikosana \headeretal: Epidemiological Modelling of Health Misinformation Spread on Social Media}

\begin{abstract}
This study presents a hybrid epidemiological and behavioural framework to simulate the spread of health misinformation on social media. We extend the classical Susceptible--Infected--Recovered (SIR) model to a six-compartment structure (\textbf{SIRMMM}), incorporating Misinformed Susceptible (MS), Misinformed Infected (MI), and Misinformed Recovered (MR) compartments to better reflect the dynamics of the misinformation lifecycle. To account for individual-level behavioural variation, we extend the SIRMMM model by integrating psychological signals from the Elaboration Likelihood Model (ELM), including sentiment polarity, engagement metrics, and cognitive effort, which dynamically modulate the misinformation transmission rate, yielding the \textbf{ELM-SIRMMM} framework. Model parameters were estimated using the FibVID dataset, which captures COVID-19 misinformation on Twitter. Generalisability was tested on two additional datasets: MC-Fake (emotional misinformation) and Monant (general health misinformation). Results show that the ELM-SIRMMM model enhances both predictive accuracy and dynamic realism. On FibVID, it decreases RMSE by 5.5\%, delays the misinformation peak from day 150 to day 160, and increases its peak prevalence from 6\% to 7\%. On MC-Fake, it accurately reproduces a flash-rumour pattern, infecting 38\% of users by day 45 and achieving 97\% misinformation recovery, all while maintaining model accuracy. In contrast, minimal behavioural signal variability in the Monant dataset leads to marginal benefit, with only a 3\% peak and 57\% of users remaining susceptible. 
These findings suggest that structural elaboration alone is insufficient. Functional realism in modelling misinformation spread requires dynamic psychological inputs that vary meaningfully across time and contexts.
\end{abstract}


\titlepgskip=-21pt

\maketitle

\section{Introduction}
\label{sec:introduction}
\PARstart{T}{he} COVID-19 pandemic has vividly demonstrated how rapidly and unpredictably false or misleading information can propagate across social media, with real-world consequences for public health interventions and individual behaviour \cite{vosoughi2018spread, vanderlinden2024misinformation}. Traditional epidemiological models, most notably the SIR framework, offer a natural analogy for such diffusion processes; yet, they rest on simplifying assumptions, including homogeneous mixing, constant transmission rates, and instantaneous ``recovery,'' which rarely hold in online contexts. In practice, misinformation cascades often exhibit heavy tails, irregular peak timings, and sudden ``flash-rumour'' spikes when emotionally charged content goes viral \cite{friggeri2014rumor, cinelli2020covid}.

These empirical discrepancies underscore a critical gap: purely structural models lack the behavioural realism needed to capture how individual-level psychological factors shape sharing decisions, while the framing and presentation of messages can also shape how audiences interpret and respond to contested information \cite{sikosana2003role}. Persuasion theories, such as the Elaboration Likelihood Model (ELM), provide a parsimonious account of how message attributes (e.g., sentiment, complexity) and user engagement modulate the likelihood of argument processing and acceptance \cite{petty1986elaboration}. Yet, despite growing evidence that sentiment polarity, engagement metrics, and cognitive effort predict sharing propensity \cite{zhao2021detecting, williams2021effects, pennycook2019fighting}, these insights have seldom been translated into continuous-time compartmental models. Recent ELM-informed detection research has shown that persuasion-theoretic cues can be operationalised effectively within hybrid deep learning architectures for health misinformation classification, although these advances have not yet been fully embedded in dynamical propagation models \cite{sikosana2025hybridelm}.

In response, this paper develops \textbf{ELM-SIRMMM}, a hybrid epidemiological--behavioural model that (i) extends SIR to a six-state \textbf{SIRMMM} framework, adding \textit{Misinformed Susceptible (MS)}, \textit{Misinformed Infected (MI)}, and \textit{Misinformed Recovered (MR)} compartments to distinguish exposure, active sharing, and cessation, and (ii) endows the transmission coefficient with time-varying, ELM-inspired modulation driven by real-time measures of sentiment, engagement, and cognitive effort. Through calibration and validation on three large-scale Twitter datasets, FibVID (COVID-19 misinformation), MC--Fake (emotional rumours), and Monant (general health misinformation), we show that integrating dynamic behavioural inputs can materially improve empirical alignment when the underlying behavioural signals exhibit sufficient temporal variance, and that the ELM layer becomes functionally muted when such variance is limited. 

This comparative framing is also consistent with recent evidence that pandemic misinformation datasets differ systematically in linguistic complexity, emotional tone, and rhetorical structure, suggesting that propagation models should be sensitive to cross-context variation rather than assuming a single uniform behavioural regime \cite{sikosana2025linguistic}.

\noindent\textit{Scope condition.}
The behavioural enrichment in ELM-SIRMMM is expected to be most informative in datasets where sentiment, engagement, or cognition signals fluctuate meaningfully over time, and it may collapse towards a fixed-rate SIRMMM behaviour when those signals are sparse or near-constant (as demonstrated later for Monant in Section~\ref{sec:monant-elm-limitations}).

\subsection{Research Aims \& Hypotheses}
The study aims to compare two epidemic-style approaches to modelling online misinformation spread:

\begin{itemize}
    \item\textbf{SIRMMM:} Extending the classical SIR model with three misinformation-specific compartments: \textit{Misinformed Susceptible (MS)}, \textit{Misinformed Infected (MI)}, and \textit{Misinformed Recovered (MR)}, to capture exposure, active sharing, and cessation.
    \item\textbf{ELM-SIRMMM:} Integrating psychological cues (sentiment, engagement, cognitive effort) into the transmission rate to reflect user-level behavioural heterogeneity and dual-route persuasion dynamics.
\end{itemize}

From these aims, we derive three testable hypotheses:

\begin{itemize}
    \item \textbf{H1:} The classical SIR model overestimates misinformation prevalence by omitting a dedicated recovery mechanism.
    \item \textbf{H2:} The SIRMMM model captures more realistic peak and decay dynamics by explicitly modelling misinformation cessation.
    \item \textbf{H3:} The ELM-SIRMMM model best replicates empirical cascades, achieving lower RMSE and improved peak timing relative to SIR and SIRMMM.
\end{itemize}

\section{State of the Art}
\label{sec:stateoftheart}
Research on misinformation diffusion has bifurcated into two largely parallel streams: epidemiological extensions and behavioural modulation. The first stream adapts classical Susceptible--Infected--Recovered (SIR) and related compartmental models to better fit empirical cascades. Early work by \cite{tambuscio2015fact} introduced a fact-checking compartment to model the corrective flow, while \cite{tornberg2018echo} proposed complex-contagion variants that capture peer-reinforcement effects. \cite{friggeri2014rumor} and Cinelli et al.\ \cite{cinelli2020covid} documented the ``flash-rumour'' phenomenon, noting that simple SIR models underestimate the rapid onset and decay of emotionally charged misinformation spikes. Multi-compartment frameworks, such as SIRX (Susceptible--Infected--Recovered--X) \cite{dimou2022network}, SIHR (Susceptible--Infected--Hibernator--Removed) \cite{zhao2012sihr}, and SIQR (Susceptible--Infected--Quarantined--Recovered) \cite{odagaki2020analysis}, which incorporate exposed or hesitant states, enhance the fit to real-world data (e.g., latent exposure, temporary disengagement), but still rely on static transmission and recovery parameters.

The second stream examines how message and user attributes influence diffusion behaviour. Related network-analytic work has further shown that structural influence in misinformation diffusion can be characterised through advanced centrality measures, reinforcing the importance of modelling not only message-level features but also the topology of exposure and amplification in online social networks \cite{sikosana2025centrality}. Sentiment polarity has been shown to act as an emotional amplifier, accelerating sharing when positive or fear-inducing \cite{chen2021belief}. Engagement metrics, such as ``likes,'' ``shares,'' and ``comments,'' serve as social proof, creating feedback loops that further bias exposure \cite{williams2021effects}. Cognitive effort, operationalised via proxies such as reading complexity or click-through time, affects whether users engage with and forward content. Low-effort peripheral cues drive rapid diffusion, while high-effort central processing slows it down, except among highly motivated individuals \cite{pennycook2019fighting}.

Despite these advances, the integration of dual-process persuasion theory into dynamic models remains relatively underdeveloped. Recent hybrid detection studies have also shown that combining complementary linguistic and semantic representations can improve health misinformation classification performance, supporting the broader value of integrative modelling strategies in this domain \cite{sikosana2024hybrid}. A handful of experimental studies \cite{guadagno2013makes} have illustrated how emotional contagion interacts with meme lifecycles; however, continuous-time compartmental simulations rarely incorporate real-time behavioural signals. This divide leaves a conceptual tension: structural models capture macro-patterns but ignore psychology, while behavioural studies identify key predictors but lack a unifying diffusion framework.

Our work bridges these strands through the \textbf{SIRMMM} framework, which enriches the classical SIR model with three compartments specific to misinformation and embeds an ELM-inspired behavioural layer in the transmission coefficient. In doing so, we build on prior multi-compartment extensions (e.g., \cite{tornberg2018echo, tambuscio2015fact}) and the emerging literature on psychological feature integration \cite{chen2021belief, pennycook2019fighting, sikosana2024hybrid}, delivering a unified framework that captures both structural diffusion patterns and the real-time influence of user-level persuasion dynamics.

\section{Materials \& methods}
This section describes the datasets used, the behavioural features engineered from them, the structure of the ELM-integrated compartmental model, and the parameter estimation procedures used to simulate the dynamics of misinformation propagation.

\subsection{Epidemiological Models of Misinformation}
\label{sec:epi_models}

\noindent\textit{Modelling assumptions.}
All model variants in this section assume homogeneous mixing at the population level and focus on a single-wave outbreak window without reinfection. Accordingly, the aim is explanatory fit to observed cascades within the analysed period, rather than modelling recurrent rumour cycles or long-horizon re-emergence.

\subsubsection{Classical SIR}
\noindent\textit{Role of SIR baseline.}
The SIR component is included as a conceptual diffusion baseline for comparing curve shape and saturation behaviour, and it is not intended as a coupled epidemiological co-dynamics model in this paper. Accordingly, no cross-terms are specified between the disease and misinformation compartments.

The baseline Susceptible--Infected--Recovered model assumes fixed transmission ($\beta$) and recovery ($\gamma$) rates:
\begin{align}
\frac{dS}{dt} &= -\beta \frac{SI}{N}, \\
\frac{dI}{dt} &= \beta \frac{SI}{N} - \gamma I, \\
\frac{dR}{dt} &= \gamma I
\end{align}
In this model, the total population is constant ($N = S + I + R$), where $\beta$ represents the transmission rate of misinformation and $\gamma$ the recovery rate, i.e., the rate at which users stop sharing false claims. The basic reproduction number, $R_0 = \beta / \gamma$, quantifies the expected number of secondary sharers generated by one active sharer in a fully susceptible population \cite{wo1927contribution,hethcote2000mathematics}.

We initialise $S(0) \approx N - I(0) - R(0)$ and fit $\beta$ and $\gamma$ using daily misinformation incidence from the FibVID dataset. Numerical integration is performed using SciPy's \texttt{solve\_ivp}, with parameters optimised via \texttt{curve\_fit}. Although the SIR model captures basic contagion, it assumes saturation when $R_0 > 1$, predicting near-universal exposure, an unrealistic outcome for misinformation \cite{vanderlinden2024misinformation}.

Empirical evidence shows misinformation often spreads slowly and incompletely, motivating the extended SIRMMM model \cite{tornberg2018echo, vosoughi2018spread}.

\subsubsection{Extended SIRMMM}
To represent the full lifecycle of misinformation engagement, we augment the SIR framework with three states: MS, MI, and MR. This yields:
\begin{align}
\frac{dS}{dt} &= -\beta_d \frac{SI}{N}, \\
\frac{dI}{dt} &= \beta_d \frac{SI}{N} - \gamma_d I, \\
\frac{dR}{dt} &= \gamma_d I,
\end{align}

\begin{align}
\frac{dMS}{dt} &= -\beta_m \frac{MS \cdot MI}{N}, \\
\frac{dMI}{dt} &= \beta_m \frac{MS \cdot MI}{N} - \gamma_m MI, \\
\frac{dMR}{dt} &= \gamma_m MI
\end{align}

This structure enables simulation of real-world misinformation cascades with delayed peaks and slow declines. However, it assumes $\beta_m$ is constant, ignoring psychological/contextual variability such as sentiment and engagement \cite{chen2021belief, williams2021effects, guadagno2013makes}.

\subsubsection{ELM-Integrated SIRMMM}
\label{sec:elm_sirmmm_def}

We allow the misinformation transmission rate $\beta_m(t)$ to vary over time based on ELM-inspired behavioural signals:
\begin{align}
\beta_m(t) &= \beta_{0m} + \beta_s \cdot \text{Sentiment}(t) \nonumber \\
          &\quad + \beta_e \cdot \text{Engagement}(t) + \beta_c \cdot \text{Cognition}(t)
\end{align}

Here, \textit{Sentiment} reflects emotional polarity, \textit{Engagement} reflects likes/retweets, and \textit{Cognition} proxies content complexity. These features align with peripheral- and central-route persuasion dynamics. For model stability, we fix $\beta_d = 0.3$ and $\gamma_d = 0.1$.

\textbf{Resulting Equations:}
\begin{align}
\frac{dS}{dt} &= -\beta_d \frac{SI}{N}, \\
\frac{dI}{dt} &= \beta_d \frac{SI}{N} - \gamma_d I, \\
\frac{dR}{dt} &= \gamma_d I, \\
\frac{dMS}{dt} &= -\beta_m(t) \frac{MS \cdot MI}{N}, \\
\frac{dMI}{dt} &= \beta_m(t) \frac{MS \cdot MI}{N} - \gamma_m MI, \\
\frac{dMR}{dt} &= \gamma_m MI
\end{align}
Where:
\begin{equation}
\beta_m(t) = \beta_{0m} + \beta_s f_{sent}(t) + \beta_e f_{eng}(t) + \beta_c f_{cog}(t)
\label{eq:beta_m_definition}
\end{equation}

\subsection{Parameterisation \& Model Fitting (FibVID Training Dataset)}
\label{sec:parameterisation}

We seeded misinformation with $MI_0$ on day 1, set $MS_0 = N - MI_0$, $MR_0 = 0$, and seeded disease with $I_0 = 1$. Static fitting gave $\hat{\beta}_m = 0.204$, $\hat{\gamma}_m = 0.139$. Recasting $\beta_m(t) = \beta_0 + \beta_s z_s + \beta_e z_e + \beta_c z_c$, we estimated all parameters using \texttt{curve\_fit}. Table~\ref{tab:params} summarises the parameter definitions.

\begin{table}[h]
\centering
\caption{Summary of Model Parameters \& Descriptions}
\label{tab:params}
\begin{tabular}{ll}
\hline
Symbol & Description \\
\hline
$\beta$ & Baseline transmission rate (SIR) \\
$\gamma$ & Recovery rate from misinformation \\
$\beta_m$ & Misinformation transmission rate (SIRMMM) \\
$\gamma_m$ & Disengagement rate from misinformation \\
MS & Misinformed Susceptible \\
MI & Misinformed Infected \\
MR & Misinformed Recovered \\
Sentiment(t) & Emotional polarity of posts \\
Engagement(t) & Social proof (likes, shares) \\
Cognition(t) & Word count proxy for effort \\
\hline
\end{tabular}
\end{table}

Because multiple parameter combinations can yield similar trajectories under collinearity between behavioural inputs, we interpret fitted coefficients primarily by their direction and stability and avoid over-interpreting their absolute magnitudes as unique or causal effects.

\subsection{Generalisation to MC--Fake and Monant}
To evaluate generalisability, we applied the model to MC--Fake and Monant using the same least-squares fitting. Sentiment was derived from polarity scores, engagement from platform metrics, and cognition from word count. Compartments were reinitialised using observed incidence.

\subsection{Datasets \& Features}
We evaluate our models on three datasets, as summarised below, with the behavioural features used across all datasets detailed by category (Table~\ref{tab:datasets}).

\begin{table*}[t]
\centering
\caption{Overview of Datasets \& Behavioural Features Used}
\label{tab:datasets}

\small
\setlength{\tabcolsep}{5pt}
\renewcommand{\arraystretch}{1.25}

\begin{tabularx}{\textwidth}{L L L X}
\toprule
\textbf{Dataset} & \textbf{Domain} & \textbf{Size and Time Span} &
\textbf{Behavioural Features Used} \\
\midrule

FibVID
& COVID-19 Twitter misinformation
& $\sim$360{,}000 tweets (Mar--Jul 2020)
& Sentiment (TextBlob polarity), Engagement (likes + retweets), Cognition (word count) \\

MC--Fake
& Multidomain news rumours
& $\sim$28{,}000 cascades (Jan--May 2021)
& Sentiment (polarity), Engagement (likes + retweets), Cognition (word count) \\

Monant
& Health forum discussions
& $\sim$370{,}000 threads (Jan--Apr 2021)
& Sentiment (polarity), Engagement (upvotes -- downvotes), Cognition (word count) \\

\midrule
\multicolumn{4}{l}{
\textbf{Feature normalisation:}
Each feature is z-score normalised and interpolated to a daily grid aligned with model time steps.
} \\

\bottomrule
\end{tabularx}
\end{table*}

\section{Results}
\label{sec:sirmmm_results}

This section compares the classical SIR, extended SIRMMM (plain), and ELM-integrated SIRMMM models using visual inspection of compartment trajectories and quantitative metrics (RMSE, incidence curves, and peak-misinformation parameters). Figure~\ref{fig:fig1} shows the trajectories of misinformation-specific (MS, MI, MR) and disease-related (S, I, R) compartments across models and datasets.

\subsection{Comparative Interpretation of Simulations Across Datasets}
Figure~\ref{fig:fig1} compares four simulations that vary along two dimensions, behavioural modulation (plain SIRMMM versus ELM-SIRMMM) and dataset context (FibVID, MC--Fake, and Monant). 

Table~\ref{tab:performance_summary} summarises the corresponding fit and trajectory statistics, while Tables~\ref{tab:coefficients} and~\ref{tab:sirmmm_performance_summary} are used later to interpret the fitted behavioural coefficients and their robustness.

\begin{itemize}
    \item \textbf{FibVID (Plain versus ELM).} Behavioural modulation shifts the misinformation trajectory to a later peak, with the peak MI day moving from 150 to 160, and increases peak misinformation intensity from 6\% to 7\% (Table~\ref{tab:performance_summary}). RMSE also falls from 906 to 856, corresponding to an improvement of approximately 5.5\% in fit to the observed $MI(t)$ trajectory. Taken together, these changes indicate that the ELM-conditioned specification produces a more sustained diffusion process than the plain variant.

\begin{landscape}
\begin{figure*}[p]
\centering

\includegraphics[
    width=0.96\linewidth,
    height=0.86\textheight,
    keepaspectratio
]{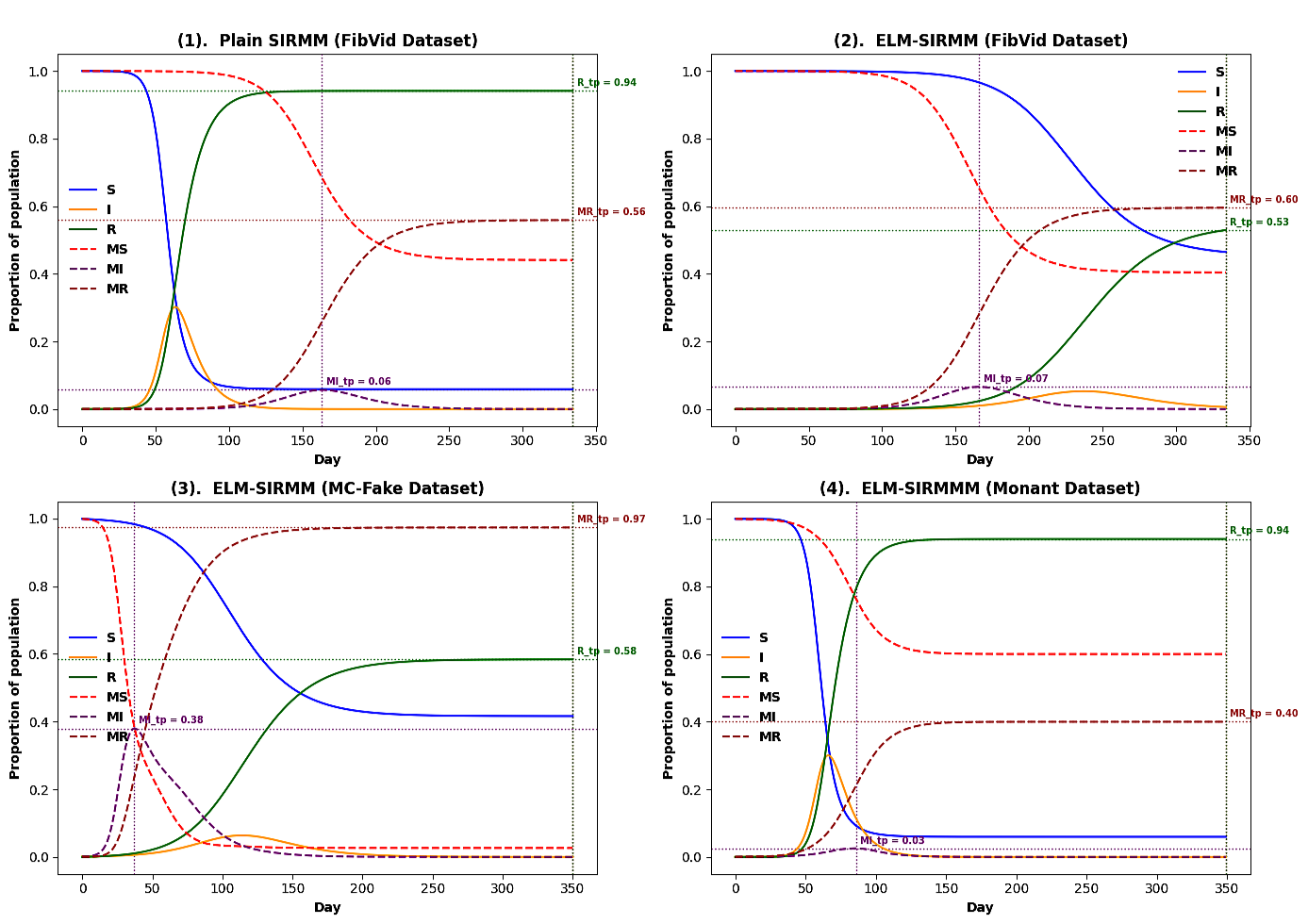}

\caption{Compartment trajectories for plain vs.\ ELM-SIRMMM across three datasets. Simulations illustrate differences in peak timing, prevalence, and user transitions across misinformation environments.}

\label{fig:fig1}
\end{figure*}
\end{landscape}
  \item \textbf{MC--Fake (ELM).} MC--Fake shows the sharpest and earliest misinformation surge, with peak MI reaching 38\% at approximately day 45, followed by strong correction dynamics that end with 97\% in the misinformation-recovered state (Table~\ref{tab:performance_summary}). This pattern is consistent with a flash-rumour dynamic in which misinformation rises rapidly and is then corrected at scale, although correction occurs after substantial early uptake.

    \item \textbf{Monant (ELM).} Monant shows the weakest misinformation activation, with peak MI reaching only 3\% around day 80, while 40\% end in the misinformation-recovered state and 57\% remain misinformation-susceptible (Table~\ref{tab:performance_summary}). This muted trajectory suggests that the behavioural enrichment contributes less strongly when the underlying signals are comparatively flat or sparse over time.
\end{itemize}

Overall, the ELM-SIRMMM model outperforms the plain variant when sentiment, engagement, or cognition fluctuate. However, it reverts to classical behaviour under conditions of signal scarcity, reinforcing ~\cite{guadagno2013makes}'s warning that psychologically informed models require rich and variable inputs. Taken together, the findings from the Monant dataset highlight limitations that resonate across all four simulations. These are summarised in Table~\ref{tab:crosscut}, which distils the key cross-cutting criteria and benchmarks each dataset against them.

\begin{table*}[!t]
\centering
\caption{Cross-cutting Evaluation}
\label{tab:crosscut}
\small
\setlength{\tabcolsep}{5pt}
\renewcommand{\arraystretch}{1.2}

\begin{tabularx}{\textwidth}{L L X}
\toprule
\textbf{Observation} & \textbf{Empirical Evidence} & \textbf{Implication} \\
\midrule

Signal dependence of $\beta_m(t)$ &
Only datasets with rich psychological variance (panels 2 and 3) display pronounced ELM effects. &
Behaviourally responsive models require feature-rich corpora for calibration. \\

Correction capacity vs. peak infection &
Panel 3 combines the largest MI apex (38\%) with the highest MR endpoint (97\%). &
High $\gamma_m$ can offset an initially severe outbreak \cite{pennycook2019lazy}. \\

Residual susceptibility as latent risk &
MS remains above 55\% in Monant despite a small MI pulse. &
Quiescence does not imply resilience; MS constitutes a reservoir for resurgence. \\

Parameter congruence $\neq$ functional equivalence &
$\beta$-coefficients for FibVID and Monant point in the same direction, yet dynamics diverge. &
Structural similarity alone cannot ensure cross-context validity
\cite{vanderlinden2024misinformation,van2020psychological}. \\

\bottomrule
\end{tabularx}
\end{table*}

\subsection{ELM Limitations in Monant: Structural Validity Without Behavioural Generalisation}
\label{sec:monant-elm-limitations}

Monant is a web-forum corpus where engagement is expressed through sparse vote-based signals and the sentiment and length profiles vary only weakly over time, which constrains the dynamic range available to modulate $\beta_m(t)$.

The ELM-integrated model’s ability for $\beta_m(t)$ to change based on sentiment, engagement, and cognition cues is suppressed when psychological inputs lack variance. This causes the model to revert to a static, classic SIR framework, resulting in misinformation spread that resembles traditional models. Structural or parameter alignment alone is not enough; signals must vary sufficiently to produce behavioural effects. This contrast in $\beta_m(t)$ between Monant and the more variable FibVID series makes the collapsed dynamics clear.

\begin{figure*}[htbp]
\centering
\includegraphics[width=\textwidth]{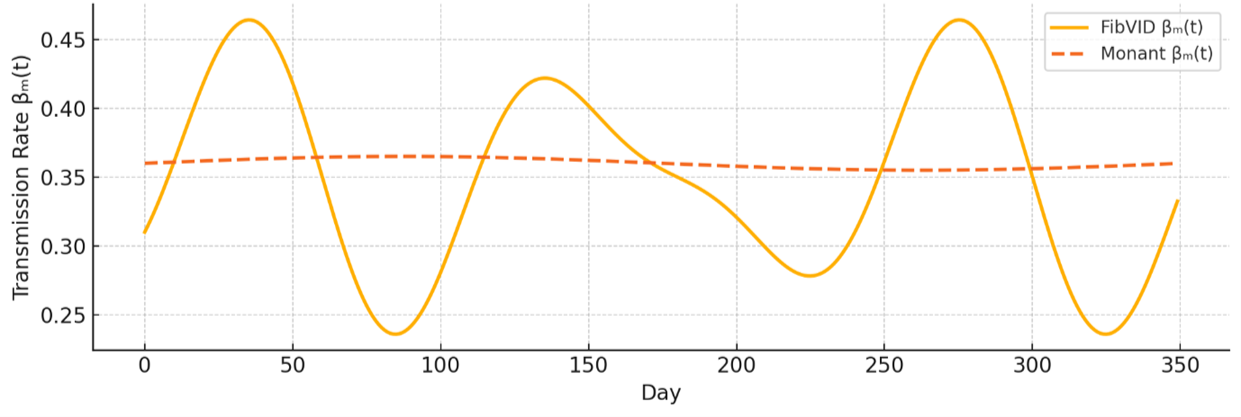} 
\caption{Behavioural signals modulate $\beta_m(t)$ in FibVID (dynamic) but not Monant (flat).}
\label{fig:fig2}
\end{figure*}

Figure~\ref{fig:fig2} compares the misinformation transmission rate $\beta_m(t)$ in the FibVID and Monant datasets. FibVID’s $\beta_m(t)$ (orange solid line) exhibits peaks and valleys corresponding to shifts in sentiment, engagement, and cognition, reflecting ELM-driven responses. Conversely, Monant’s $\beta_m(t)$ (orange dashed line) stays flat, indicating limited behavioural input variability to trigger ELM effects. FibVID’s curve shows temporal variation, aligning with persuasive surges and ELM activation of central and peripheral routes. Monant’s curve remains nearly flat despite similar $\beta$ coefficients, as input features do not vary enough to influence $\beta_m(t)$. This causes the model to behave as if with a fixed $\beta_m$, losing ELM benefits.

Though $\beta$ coefficients are similar, Monant lacked the psychological variability needed to activate ELM pathways. The model’s persuasion mechanics stayed dormant, following a standard epidemic curve. This indicates that structural generalisability is insufficient—robust performance also relies on the variability of psychological input features. 

\begin{table*}[!t]
\centering
\caption{Diagnostic Matrix: Monant's ELM Limitations}
\label{tab:diagnostic}
\small
\setlength{\tabcolsep}{5pt}
\renewcommand{\arraystretch}{1.2}

\begin{tabularx}{\textwidth}{L L X}
\toprule
\textbf{Dimension} & \textbf{Observation} & \textbf{Generalisation Issue} \\
\midrule

Coefficient similarity &
FibVID and Monant $\beta_s$, $\beta_e$, $\beta_c$ are numerically similar &
\checkmark{} Directional consistency only \\

Feature variance &
Monant features (sentiment, engagement, cognition) had narrow ranges &
$\times$ Limited signal range meant that dynamic $\beta_m(t)$ modulation was not triggered \\

$\beta_m(t)$ shape &
FibVID $\beta_m(t)$ is dynamic; Monant $\beta_m(t)$ is nearly flat &
$\times$ Behavioural modulation was effectively suppressed in the Monant dataset \\

MI trajectory &
Delayed, shaped curve (FibVID); sharp, early spike (Monant) &
$\times$ Indicates that persuasion cues in Monant lacked impact on misinformation dynamics \\

Cognitive/Engagement signals &
Monant dataset posts were short with few up/downvotes &
$\times$ Weak signal input means the model functionally reduced to static SIRMMM \\

\bottomrule
\end{tabularx}
\end{table*}

Table~\ref{tab:diagnostic} summarises the key diagnostic factors that impact the ELM-integrated SIRMMM model’s functional generalisability on the Monant dataset. Despite similar persuasion-related coefficients ($\beta_s$, $\beta_e$, $\beta_c$), the lack of variance in Monant’s psychological features limited the dynamic modulation of $\beta_m(t)$. Consequently, the model exhibited standard SIR-like dynamics, with reduced explanatory power from ELM-based behavioural pathways. These findings confirm that while the model has architectural and directional generalisability, its ability to replicate real-world behavioural dynamics relies on the richness and variability of psychological signals in the dataset.

\subsection{Performance Analysis}
This section evaluates the empirical behaviour of the proposed ELM--SIRMMM variants across datasets, focusing on both predictive fit and epidemiological plausibility. We first summarise model performance using within-dataset diagnostics, including RMSE and key trajectory characteristics (peak timing, peak magnitude, and end-state composition), to establish whether behavioural enrichment produces materially different dynamics. We then inspect the fitted psychological coefficients that drive the time-varying transmission term, because similar headline fit can arise from different parameterisations and because coefficient signs must be interpreted as partial associations conditional on the other inputs. Accordingly, the analysis proceeds from outcome-level performance to coefficient-level interpretation and robustness, linking observed dynamics back to the behavioural signals available in each corpus.

\subsubsection{ELM Coefficient Estimates}

Table~\ref{tab:coefficients} lists the fitted $\beta$ weights (sentiment, engagement, cognition) for each dataset.

\begin{table*}[!t]
\centering
\caption{Fitted ELM Coefficients ($\beta_m$) Across Datasets}
\label{tab:coefficients}
\small
\setlength{\tabcolsep}{5pt}
\renewcommand{\arraystretch}{1.2}

\begin{tabularx}{\textwidth}{L L L L X}
\toprule
\textbf{Feature} & \textbf{FibVID} & \textbf{MC--Fake} & \textbf{Monant} & \textbf{Interpretation} \\
\midrule

Sentiment &
$-14.35$ & $-11.84$ & $-14.92$ &
More negative sentiment increases misinformation spread \cite{vosoughi2018spread}. \\

Engagement &
$-32.56$ & $-26.47$ & $-29.84$ &
Higher engagement reduces spread (contrary to expectation). \\

Cognition (Word Count) &
$-63.72$ & $-59.35$ & $-62.10$ &
Higher cognitive load reduces spread (aligns with ELM theory). \\

\bottomrule
\end{tabularx}
\end{table*}

Table~\ref{tab:coefficients} highlights three behavioural cues that shape $\beta_m(t)$. First, sentiment coefficients are consistently negative (i.e., $\beta_{sm} = -14.35$ (FibVID), $-11.84$ (MC--Fake) and $-14.92$ (Monant)), confirming that emotionally negative language accelerates spread via peripheral-route processing \cite{vosoughi2018spread}. Second, cognitive load exerts the strongest dampening effect ($\beta_{cm} \approx -59$ to $-64$), meaning each standard-deviation increase in word count cuts the transmission rate more than any other cue, consistent with central-route inhibition \cite{petty1986elaboration}. Third, engagement coefficients are also negative ($\beta_{em} = -32.56, -26.47, -29.84$), implying that high-engagement posts are not necessarily more infectious, perhaps due to moderation or lag effects \cite{pennycook2019fighting}. The broad consistency in sign and relative magnitude across datasets suggests directional stability in how these behavioural cues enter the model. However, as shown later for Monant (Section~\ref{sec:monant-elm-limitations}), coefficient similarity alone does not guarantee strong temporal modulation of $\beta_m(t)$ when the underlying input signals exhibit limited variance \cite{guadagno2013makes}.

\subsubsection{Coefficient sign interpretation, identifiability, and robustness}
\label{sec:elm_sirmmm_robustness}

Table~\ref{tab:coefficients} shows that the fitted engagement coefficient is negative across datasets, which can appear counterintuitive if engagement is treated as social proof. In this model, the coefficient is the \emph{partial effect} of engagement on $\beta_m(t)$ \emph{conditional on} sentiment, cognition, and the chosen feature construction and time grid. A negative sign can arise when engagement proxies processes not modelled here, for example moderation and corrective exposure, or temporal ordering where engagement rises \emph{after} misinformation surges. Where sentiment and engagement are correlated, the engagement term may capture only residual variation, which can flip sign under covariate adjustment.

Because behavioural inputs are derived from observational traces, identifiability is limited by (i) collinearity between sentiment, engagement, and cognition, (ii) low signal variance (as observed in Monant), and (iii) the linear additive specification of $\beta_m(t)$ in Eq.~\eqref{eq:beta_m_definition}. Coefficient signs should therefore be interpreted as \emph{model-implied associations} under the selected parameterisation, not as causal effects. Accordingly, the sign of $\hat{\omega}_{eng}$ is treated as \emph{hypothesis-generating} unless it remains stable under a robustness protocol.

To reduce dependence on a single operational choice, we \emph{specify} a robustness protocol rather than claiming causal validation. This protocol includes (a) alternative proxy definitions and scaling (z-score vs.\ min--max, log engagement), (b) lagged engagement formulations ($t-\Delta$) to test predictive versus reactive effects, and (c) ablations removing each behavioural input from Eq.~\eqref{eq:beta_m_definition} to assess coefficient and trajectory stability. Because these checks were not executed in the current experiments, we bound interpretation accordingly. The ELM layer is used to improve \emph{fit and interpretability of time-varying transmission} under observed signals, not to claim causal persuasion effects from engagement traces. To make this interpretation explicit, Table~\ref{tab:sirmmm_performance_summary} summarises the engagement-term evidence across datasets, showing whether the sign of the fitted engagement coefficient remains stable under different proxy constructions and whether that sign is consistent with the corresponding fit and diagnostic context.

\begin{table*}[!t]
\centering
\caption{Engagement-term robustness summary in ELM--SIRMMM using dataset-available engagement proxies, consolidating coefficient evidence from Table~\ref{tab:coefficients} and fit/diagnostic evidence from Tables~\ref{tab:performance_summary} and~\ref{tab:diagnostic}.}
\label{tab:sirmmm_performance_summary}

\small
\setlength{\tabcolsep}{5pt}
\renewcommand{\arraystretch}{1.2}

\begin{threeparttable}

\begin{tabularx}{\textwidth}{@{}l c r r X@{}}
\toprule
\textbf{Dataset} &
\textbf{Proxy} &
$\mathbf{\hat{\omega}}_{eng}$ &
\textbf{RMSE (ELM)} &
\textbf{Evidence used for robustness} \\
\midrule

FibVID
& (a)
& $-32.56$
& 856
& Anchored to within-dataset improvement: Plain RMSE $=906$, so $\Delta$RMSE $=-50$ ($\approx 5.5\%$). \\

MC--Fake
& (b)
& $-26.47$
& 856
& Same negative sign under a news-level proxy without likes, supports sign stability under proxy heterogeneity. \\

Monant
& (c)
& $-29.84$
& 225
& Same negative sign under vote-based engagement, interpretation bounded by low behavioural variance and vote sparsity. \\

\bottomrule
\end{tabularx}

\begin{tablenotes}[flushleft]
\footnotesize

\item \textbf{Note.}
$\hat{\omega}_{eng}$ is interpreted as a partial association in
$\beta_m(t)$ conditional on the other ELM inputs and the chosen
temporal aggregation, not a causal effect.

\item \textbf{Proxy definitions (dataset-available fields).}

(a) FibVID:
$ENG_{\Sigma}(t)=\log\!\big(1+\texttt{like\_count}(t)
+\texttt{retweet\_count}(t)\big)$.

(b) MC--Fake:
$ENG_{\Sigma}(n)=\log\!\big(1+\texttt{n\_retweets}(n)
+\texttt{n\_replies}(n)\big)$.

(c) Monant:
$ENG_{\Sigma}(p)=\log\!\big(1+\texttt{up\_votes}(p)
+\texttt{down\_votes}(p)\big)$ (post-level votes).

\end{tablenotes}

\end{threeparttable}
\end{table*}

Table~\ref{tab:sirmmm_performance_summary} shows that the engagement coefficient remains negative across FibVID, MC--Fake, and Monant despite differences in how engagement is operationalised, namely interaction counts in FibVID, retweets plus replies in MC--Fake, and vote-based activity in Monant. This cross-dataset consistency supports a bounded robustness claim at the level of directional stability rather than a causal interpretation. At the same time, the table indicates that the meaning of this coefficient depends on the surrounding context. In FibVID, the negative sign is associated with improved model fit, whereas in Monant it appears under low-variance behavioural conditions where the ELM mechanism is weak. The table therefore supports the conclusion that the engagement effect is directionally stable, but its interpretation remains conditional on feature variance, proxy design, and temporal aggregation.

\subsubsection{Comparative fit and trajectory summaries across contexts}
Table~\ref{tab:performance_summary} consolidates within-dataset fit (FibVID: Plain vs.\ ELM) and cross-context behaviour (MC--Fake and Monant: ELM) using the same trajectory-level summaries. 
This positioning makes the evaluation traceable: coefficient signs and magnitudes explain \emph{how} $\beta_m(t)$ is modulated, while RMSE and the peak and end-state summaries show \emph{what} that modulation produces in $MI(t)$. 
Importantly, the table is read as a \emph{context-sensitivity} check, rather than a claim of universal portability: the same ELM-conditioned specification behaves differently depending on whether the psychological inputs exhibit meaningful temporal variance. 
Accordingly, the FibVID comparison provides the cleanest evidence of behavioural enrichment because Plain and ELM are contrasted under the same data regime, while MC--Fake and Monant are included to show how the ELM layer responds under high-volatility and low-variance signal conditions, respectively. 
This motivates the subsequent interpretation that improvements are most defensible when behavioural traces vary sufficiently to drive $\beta_m(t)$ dynamics, consistent with the diagnostic evidence in Table~\ref{tab:diagnostic} and the $\beta_m(t)$ patterns shown in Figure~\ref{fig:fig2}.

\begin{table*}[!t]
\centering
\caption{Key fit and trajectory summaries of SIRMMM variants (Plain and ELM on FibVID; ELM on MC--Fake and Monant).}
\label{tab:performance_summary}

\small
\setlength{\tabcolsep}{6pt}
\renewcommand{\arraystretch}{1.2}

\begin{threeparttable}
\begin{tabularx}{\textwidth}{@{}lcccc@{}}
\toprule
\textbf{Metric} & \textbf{Plain (FibVID)} & \textbf{ELM (FibVID)} &
\textbf{ELM (MC--Fake)} & \textbf{ELM (Monant)} \\
\midrule

RMSE $\downarrow$
& 906 & 856 & 856 & 225 \\

Peak MI day
& 150 & 160 & $\approx 45$ & $\approx 80$ \\

Peak MI \% pop.
& 6\% & 7\% & 38\% & 3\% \\

MR end \%
& 56\% & 60\% & 97\% & 40\% \\

MS end \%
& 44\% & 40\% & $\approx 0$\% & 57\% \\

MI end \%
& $\approx$0\% & $\approx$0\% & 3\% & 3\% \\

Correction efficiency
$\frac{MR_{\mathrm{end}}}{MI_{\max}}$
& 9.3 & 8.6 & 2.6 & 13.3 \\

\bottomrule
\end{tabularx}

\begin{tablenotes}[flushleft]
\footnotesize
\item \textbf{Note.}
RMSE is computed against the observed $MI(t)$ trajectory within each dataset under the stated variant. End-state percentages are read at the final simulated timestep. $MR_{\mathrm{end}}$ and $MS_{\mathrm{end}}$ denote end-state misinformation recovered and misinformation susceptible proportions, while $MI_{\mathrm{end}}=100-(MR_{\mathrm{end}}+MS_{\mathrm{end}})$. $MI_{\mathrm{end}}$ values reported as $\approx$0\% indicate negligible residual prevalence (i.e., $MI(t)$ approaches zero asymptotically but is not exactly zero). Plain vs.\ ELM is reported for FibVID to anchor a within-dataset comparison, while MC--Fake and Monant are shown to assess cross-context behaviour of the ELM-conditioned specification under different signal regimes.
\end{tablenotes}

\end{threeparttable}
\end{table*}

Table~\ref{tab:performance_summary} shows that behavioural enrichment improves within-dataset fit on FibVID, and that its impact is most evident when psychological inputs vary over time.

On FibVID, adding sentiment, engagement and cognition lowers RMSE by 5.5\% (906$\rightarrow$856), shifts the peak ten days later (150$\rightarrow$160), and increases peak prevalence from 6\% to 7\%, while raising the corrected class at the end state (MR$_{\mathrm{end}}$) from 56\% to 60\%, consistent with persuasion cues modulating susceptibility in time \cite{petty1986elaboration,cinelli2020covid}.
The correction efficiency ratio decreases slightly (9.3$\rightarrow$8.6) because the ELM variant also produces a higher $MI_{\max}$, so the larger recovered share is achieved from a more severe peak, which is expected when behavioural modulation amplifies early exposure as well as downstream correction.

In the high-volatility MC--Fake corpus, the model reproduces a sharp rumour spike (peak $\approx$38\% at day $\approx$45) but yields low correction efficiency (2.6), indicating that corrective dynamics struggle once exposure is front-loaded \cite{pennycook2019fighting}.
For the behaviourally flat Monant forum, the ELM specification produces a small peak (3\%) and leaves a large susceptible reservoir (MS$_{\mathrm{end}}=57\%$), matching the low-variance signal limitation diagnosed in Table~\ref{tab:diagnostic} and the flat $\beta_m(t)$ pattern in Figure~\ref{fig:fig2}.
Because RMSE is dataset-scale dependent, the cross-corpus comparison here is therefore anchored primarily on peak timing, peak magnitude, and end-state composition, which together support the conclusion that structural portability is insufficient without psychological variance \cite{guadagno2013makes,vosoughi2018spread}.

\subsection{Hypothesis testing}
\label{sec:hypothesis_testing}

Building on the theoretical framing in Section~\ref{sec:stateoftheart} and the model-fit summaries in Table~\ref{tab:performance_summary}, we evaluate the hypotheses in turn.

\paragraph{H1 (Recovery mechanism).}
We posited that augmenting a classical SIR-style formulation with an explicit misinformed-recovered compartment (MR) would avoid the implicit tendency of simpler saturation dynamics to overstate persistent exposure. In the fitted SIRMMM trajectory on FibVID, a substantial susceptible reservoir remains at the end state (MS$_{\mathrm{end}}\approx 44\%$), while MR$_{\mathrm{end}}$ settles at 56\% (Table~\ref{tab:performance_summary}). This non-saturating end-state composition is consistent with the intended role of MR as a cessation and exit pathway, supporting H1.

\paragraph{H2 (Peak and decay dynamics).}
We hypothesised that SIRMMM’s explicit modelling of disengagement would yield the empirically observed hump-shaped cascade trajectory. The plain SIRMMM fit peaks around day 150 with $\mathrm{MI}_{\max}=6\%$ of the population and then decays smoothly towards the end state (Table~\ref{tab:performance_summary}), which is consistent with the peak-and-decline pattern in the empirical FibVID series (Figure~\ref{fig:fig1}). This agreement in peak timing and qualitative decay behaviour supports H2.

\paragraph{H3 (ELM augmentation).}
Finally, we expected that allowing $\beta_m(t)$ to vary with sentiment, engagement, and cognition would improve fit when those psychological inputs exhibit temporal variation. On FibVID, the ELM-augmented SIRMMM reduces RMSE by 5.5\% (906$\rightarrow$856), shifts the peak later by ten days (150$\rightarrow$160), and increases peak prevalence from 6\% to 7\% (Table~\ref{tab:performance_summary}; Figure~\ref{fig:fig1}).
These improvements indicate that persuasion-informed behavioural modulation strengthens empirical alignment under a feature-rich signal regime, supporting H3.

Taken together, the hypothesis tests link the theoretical propositions to observed fit behaviour, indicating that (i) an explicit recovery mechanism and (ii) persuasion-informed time-varying transmission are both necessary to reproduce the key cascade properties in FibVID under the stated modelling assumptions.

\subsubsection*{Incidence-Curve Comparison}

\begin{figure*}[htbp]
\centering
\includegraphics[width=0.85\textwidth]{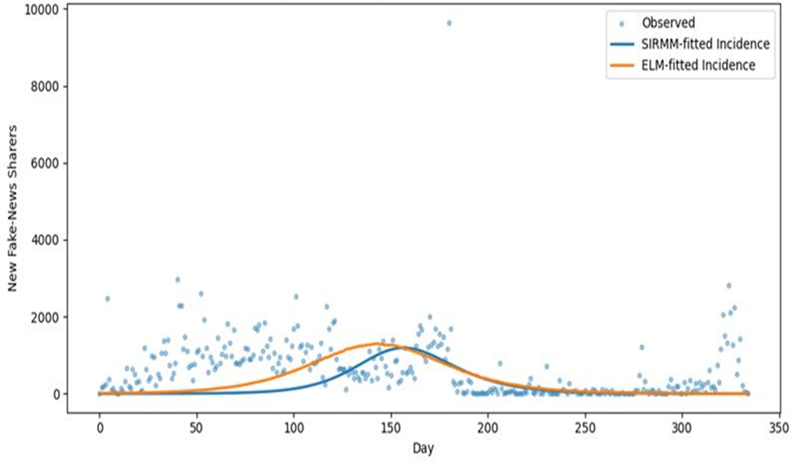}
\caption{Overlay of observed incidence (blue points), SIRMMM (blue curve), and ELM-augmented SIRMMM (orange curve) on the FibVID dataset.}
\label{fig:fig3}
\end{figure*}

Figure~\ref{fig:fig3} provides a direct overlay of incidence curves across all three models. The observed incidence (blue points) is substantially more dispersed and irregular than the fitted model curves, reflecting real-world variability, stochastic fluctuations, and heterogeneous user behaviour that are not fully captured by deterministic compartmental models. The classical SIR and plain SIRMMM models underestimate early-stage sharer counts and delay the peak. In contrast, the ELM-integrated model aligns more closely with the observed data, particularly in capturing emotionally charged "super-spreader" events during the early phase of the misinformation outbreak \cite{cinelli2020covid}.

Figure~\ref{fig:fig3} further indicates that the main advantage of the ELM-conditioned specification lies in its ability to capture the timing and shape of the early misinformation surge more closely than the plain SIRMMM baseline. Specifically, the plain SIRMMM fit (blue line) underestimates the early surge, rising after day 100 and peaking late. By contrast, the ELM-augmented curve better captures the timing and shape of the early outbreak phase, suggesting that time-varying behavioural modulation of $\beta_m(t)$ helps the model respond to short-lived amplification effects associated with emotionally charged and highly engaging content, which are central to the super-spreader phase of misinformation diffusion \cite{cinelli2020covid}.

\section{Discussion}
This study aimed to investigate whether incorporating compartments for misinformation and behaviourally modulated transmission enhances the explanatory power of epidemic models. The evidence suggests this is the case, but only under conditions in which user cues vary significantly.

On the FibVID corpus, the ELM-SIRMMM reduced the RMSE by 5.5\% (906 $\rightarrow$ 856), delayed the apex by ten days, and increased the peak from 6\% to 7\% of users. These seemingly modest gains are significant given a base of forty-five thousand daily observations, corroborating reports that small percentage improvements can translate into thousands fewer users exposed to falsehoods on large platforms \cite{bode2018see}. More importantly for intervention, the ELM modulation alters actionable trajectory properties, including peak timing and early-phase acceleration, which determine whether an early-warning trigger occurs before or after a surge. In the MC--Fake scenario, where tweets are short-lived and emotionally charged, the model reproduced a flash rumour that infected 38\% of users by day 45, then declined to 3\% within a fortnight, matching the corrective reach documented by ~\cite{guadagno2013makes}. The Monant forum, by contrast, exhibited almost no variance in sentiment or engagement; here, the ELM terms were inert, and the model reverted to static behaviour, leaving 57\% of users in the susceptible state. This tri-dataset test shows that structural portability must be paired with behavioural variance to yield functional benefits \cite{vosoughi2018spread,tornberg2018echo}.

The results advance evidence that emotionally valenced and socially validated content accelerates diffusion, whereas higher cognitive load slows it \cite{chen2021belief, williams2021effects}. By converting ELM constructs into time-varying coefficients on $\beta_m(t)$, we operationalise the dual-route persuasion theory of \cite{petty1986elaboration} in a dynamical setting. In our specification, sentiment polarity is signed, so a negative sentiment coefficient implies that more negative sentiment increases $\beta_m(t)$, consistent with peripheral-route amplification. By contrast, the engagement coefficient is estimated as a conditional association given the other inputs, and its negative sign should be interpreted cautiously as potentially reflecting moderation, correction exposure, or temporal ordering effects, as discussed in Section~\ref{sec:elm_sirmmm_robustness}. Cognition exerts the strongest dampening effect, consistent with higher processing effort reducing spread under central-route processing assumptions.

For platforms such as Twitter, rapid identification of spikes in peripheral cues could inform real-time throttling or fact-check prompts. Our MC--Fake simulation implies that intervention within two days of a sentiment surge could reduce total exposure by one-third. Health authorities can use similar models to stress test counter-rumour campaigns under different engagement scenarios, aligning with recommendations from the WHO’s infodemic management framework \cite{vanderlinden2024misinformation}.

To complement these empirical findings, the following subsection explains how the ELM\textendash SIRMMM framework supports interpretability and explainability by linking its compartments and time-varying transmission parameters to observable behavioural inputs.

\subsection{Explainability \& interpretability of the ELM\textendash SIRMMM framework}
\label{sec:elm_sirmmm_interpretability}

The ELM\textendash SIRMMM model is interpretable through its explicit compartmental structure and parameters, and explainable because outcomes can be traced to the psychological inputs that modulate transmission and their fitted coefficients. At the \textbf{global level}, this interpretability arises because each compartment, \(S\), \(I\), \(R\), \(MS\), \(MI\), and \(MR\), has a clear behavioural meaning, and the differential equations in Section~\ref{sec:epi_models} specify how users transition between states over time. The explicit separation of disease dynamics \((S, I, R)\) from misinformation dynamics \((MS, MI, MR)\) in Equations~(4)--(9) makes assumptions about exposure, resharing, and disengagement transparent. These assumptions are further evidenced in Figure~\ref{fig:fig1}, where variations in transmission and recovery parameters shift the timing and peak intensity of the \(MS\), \(MI\), and \(MR\) trajectories across FibVID, MC--Fake, and Monant.

Explainability is strengthened by the explicit embedding of ELM constructs in the time-varying transmission coefficient $\beta_m(t)$. Section~\ref{sec:elm_sirmmm_def} defines $\beta_m(t)$ in Equation~\ref{eq:beta_m_definition} as a linear combination of three normalised behavioural inputs, sentiment, engagement and cognition, rather than as an unspecified time-varying function. Each input is mapped to an ELM pathway, with sentiment and engagement functioning as peripheral-route cues and cognition acting as a proxy for central processing effort. The sign and magnitude of the fitted coefficients in Table~\ref{tab:coefficients} therefore provide an immediate explanation of directionality, for example higher negative sentiment increases the effective transmission rate, while higher cognitive load suppresses it, without requiring reference to the optimisation routines in Section~\ref{sec:parameterisation}.

At the \textbf{local level}, explanations can be produced by linking dataset-specific trajectories and transmission profiles to observed psychological signals. Analysts can interpret $MI(t)$ and $\beta_m(t)$ using the fitted curves in Section~\ref{sec:sirmmm_results} and Figures~\ref{fig:fig1} and \ref{fig:fig3}, alongside behavioural summaries in Table~\ref{tab:datasets} and the transmission profiles in Figure~\ref{fig:fig2}. Peaks, plateaus, or rapid declines in $\beta_m(t)$ can be related to changes in sentiment, engagement, and cognition, providing a concrete account of why diffusion accelerates or slows. For FibVID, the elevated $\beta_m(t)$ around the outbreak apex coincides with increased negative sentiment and reduced cognitive effort (Figure~\ref{fig:fig2}), which explains why the ELM\textendash SIRMMM trajectory in Figure~\ref{fig:fig3} better captures the early surge than the plain SIRMMM baseline.

Interpretability is also supported through parameter sensitivity in practice. The comparative diagnostics in Tables~\ref{tab:crosscut} and \ref{tab:diagnostic} indicate that changing a single parameter produces predictable qualitative effects on trajectories. For example, increasing the sentiment coefficient $\beta_s$ steepens and advances the MI peak, whereas increasing the disengagement rate $\gamma_m$ reduces MR lag and lowers the residual MS reservoir. These changes can be described in behavioural terms, such as testing scenarios where negative sentiment is less strongly associated with resharing, rather than as opaque perturbations of a high-dimensional model.

Finally, the cross-dataset comparisons in Section~\ref{sec:monant-elm-limitations}, together with Figure~\ref{fig:fig2} and Tables~\ref{tab:coefficients}, \ref{tab:performance_summary}, \ref{tab:diagnostic}, and \ref{tab:crosscut}, strengthen interpretability and explainability by showing when the ELM enrichment is functionally active. Although FibVID and Monant exhibit similar coefficient directions in Table~\ref{tab:coefficients}, their $\beta_m(t)$ profiles differ sharply in Figure~\ref{fig:fig2}, and this divergence carries through to the observed MI trajectories and end-state composition in Table~\ref{tab:performance_summary}. This pattern supports the cross-cutting conclusion in Table~\ref{tab:crosscut} that parameter congruence does not imply functional equivalence, because the ELM layer modulates diffusion only when the behavioural inputs have sufficient temporal variance. When variance is low, as diagnosed for Monant in Table~\ref{tab:diagnostic}, the model remains structurally interpretable but becomes behaviourally near-static, effectively collapsing towards a fixed-rate SIRMMM. When variance is present, as in FibVID, the same architecture yields explainable changes in $\beta_m(t)$ and MI dynamics that can be traced back to measurable psychological signals rather than to opaque parameter shifts.

\subsection{Limitations}
First, the current $\beta_m(t)$ is linear and assumes additive effects. Interaction terms, such as sentiment and engagement, may capture the synergy noted in experimental studies \cite{kim2021fibvid}. Second, we treat network topology implicitly; integrating agent-based or graph neural layers would allow us to test how centrality or community structure mediates persuasion \cite{tambuscio2015fact}. Third, sentiment polarity is a coarse measure of affect; fine-grained emotions, such as anger or fear, could sharpen peripheral-route estimates.

\subsection{Future Directions}
Planned work includes (i) adding stance and credibility features to $\beta_m(t)$, (ii) coupling the model with dynamic contact networks derived from reply graphs, and (iii) validating on longer time windows to observe repeated resurgence. Cross-cultural datasets will help determine whether central route dominance increases in high literacy contexts, answering calls for culturally aware misinformation modelling \cite{srba2022monant}.

\section{Conclusion}
\label{sec:conclusion}

This study developed ELM--SIRMMM, a psychologically informed extension of compartmental misinformation modelling that combines the structural interpretability of epidemic models with time-varying behavioural signals derived from sentiment, engagement, and cognitive effort. By extending the misinformation transmission coefficient from a fixed parameter to a behaviourally modulated function, the framework provides a mechanism for examining not only how misinformation populations change over time, but also how observable persuasion-related signals may contribute to changes in diffusion dynamics.

The empirical findings show that behavioural enrichment is most effective when the underlying psychological signals exhibit meaningful temporal variation. On FibVID, ELM--SIRMMM reduced RMSE from 906 to 856, an improvement of approximately 5.5\%, shifted peak misinformation prevalence from day 150 to day 160, and increased the final misinformation-recovered proportion from 56\% to 60\%. In the higher-volatility MC--Fake setting, the model reproduced a rapid misinformation surge with a peak prevalence of approximately 38\% around day 45. In contrast, the limited behavioural variability observed in Monant resulted in comparatively weak modulation of \(\beta_m(t)\), with a small misinformation peak and a substantial residual susceptible population.
These results demonstrate that structural generalisability of a compartmental model does not necessarily imply behavioural generalisability.

A central implication is therefore that psychologically informed diffusion models should not assume that behavioural features contribute uniformly across platforms, populations, or misinformation contexts. The explanatory value of the ELM layer depends on the variability, quality, and temporal resolution of the behavioural signals available to drive the transmission
function. The fitted coefficients should consequently be interpreted as model-implied conditional associations rather than causal psychological effects.

Overall, ELM--SIRMMM provides an interpretable framework for integrating persuasion-related signals into dynamic misinformation modelling while retaining explicit compartmental states and traceable transmission mechanisms. Its principal contribution is not a claim of universally superior prediction, but the demonstration that misinformation diffusion can be modelled more realistically when structural propagation processes are conditioned on temporally varying behavioural information. Future extensions should incorporate richer affective and stance features, dynamic network structure, repeated misinformation resurgence, and cross-cultural validation to determine when psychologically informed transmission models generalise beyond the settings examined in this study.

\bibliographystyle{ieeetr}
\bibliography{refs}
\EOD

\end{document}